\documentclass[twocolumn,preprintnumbers,floats,prd,amssymb,floatfix,nofootinbib,balancelastpage,superscriptaddress,amsmath]{revtex4-1}

\pdfoutput=1
\usepackage[utf8]{inputenc}
\usepackage{amssymb}
\usepackage{amsmath}
\usepackage{amsfonts}
\usepackage{graphicx}
\usepackage{color}
\usepackage{xspace}
\usepackage{comment}
\usepackage{hyperref}
\usepackage[normalem]{ulem}

\usepackage[section]{placeins}
\usepackage{afterpage}

\usepackage{float}
\usepackage{slashed}
\usepackage{ulem}
\usepackage{verbatim}

\usepackage{multirow,rotating}
\usepackage[dvipsnames]{xcolor}

\def\gsim{\raise0.3ex\hbox{$\;>$\kern-0.75em\raise-1.1ex\hbox{$\sim\;$}}}
\def\lsim{\raise0.3ex\hbox{$\;<$\kern-0.75em\raise-1.1ex\hbox{$\sim\;$}}}

\newcommand{\ba}[1]{\begin{eqnarray} \label{(#1)}}
\newcommand{\ea}{\end{eqnarray}}

\definecolor{dcolour}{rgb}{.5, .5, .5}

\def\gsim{\raise0.3ex\hbox{$\;>$\kern-0.75em\raise-1.1ex\hbox{$\sim\;$}}}
\def\lsim{\raise0.3ex\hbox{$\;<$\kern-0.75em\raise-1.1ex\hbox{$\sim\;$}}}

\newcommand{\iab}{\rm ab^{-1}}

\newcommand{\iTeV}{\rm TeV^{-1}}

\newcommand{\neutralino}[1]{\tilde{\chi}_{#1}^0}

\newcommand{\fb}{\,{\mathrm{fb}}}
\newcommand{\GeV}{\,{\mathrm{GeV}}}
\newcommand{\TeV}{\,{\mathrm{TeV}}}

\begin{document}

\title{Search for long-lived axions with far detectors at future lepton colliders }

\author{Minglun Tian}
\email{minglun.tian@whut.edu.cn}
\affiliation{Department of Physics, School of Science, Wuhan University of Technology, 430070 Wuhan, Hubei, China }

\author{Zeren Simon Wang}
\email{wzs@mx.nthu.edu.tw}
\affiliation{Department of Physics, National Tsing Hua University, Hsinchu 300, Taiwan}
\affiliation{Center for Theory and Computation, National Tsing Hua University, Hsinchu 300, Taiwan}

\author{Kechen Wang}
\email{kechen.wang@whut.edu.cn (Corresponding author)}
\affiliation{Department of Physics, School of Science, Wuhan University of Technology, 430070 Wuhan, Hubei, China }


\begin{abstract}
In our previous work [Phys. Rev. D 101 (2020) 075046], we have proposed to install FAr Detectors at the Electron Positron Collider (FADEPC) to enhance the discovery potential of long-lived particles (LLPs). In this study, we consider eight designs of far detectors with different locations, volumes and geometries and investigate their potential for discovering long-lived axion-like particles (ALPs) via the process  $e^-e^+ \rightarrow  \gamma \,\, a,~ a \to \gamma\gamma $  at future $e^{-}e^{+}$ colliders running at a center-of-mass energy of 91.2 GeV and integrated luminosities of 16, 150, and 750 ab$^{-1}$. We estimate their sensitivities on the model parameters in terms of the effective ALP-photon-photon coupling $C_{\gamma \gamma} / \Lambda $, the effective ALP-photon-$Z$ coupling $C_{\gamma Z} / \Lambda$, and ALP mass $m_a$ for three physics scenarios:  $C_{\gamma Z} = 0$; $C_{\gamma Z} = C_{\gamma \gamma}$ and both $C_{\gamma Z}$ and $C_{\gamma \gamma}$ can freely change.
The results provide references for the optimization of far detectors at future electron-positron colliders.
\end{abstract}


\vskip10mm

\maketitle
\flushbottom
%
%

\section{Introduction}
\label{sec:intro}
The research interests in new particles with a relatively long lifetime, the long-lived particles (LLPs), have been growing rapidly, cf. reviews~\cite{Alekhin_2016, Lee:2018pag,Alimena:2019zri,Alimena:2021mdu} and references therein for recent studies.
At colliders, such LLPs are usually produced at the interaction point (IP), travel a macroscopic distance and decay into standard model (SM) and/or other new particles.
If their lifetime is long, they have more probability of travelling a long distance and decaying outside the detectors. Because LLPs usually have feeble couplings to detector materials, if they have a neutral charge, they go out of the detector undetected and, hence, their energies and momenta are manifested as missing energy.
On the other hand, if their lifetime appropriately matches the detector size, LLPs have more probability to decay inside the detector, giving more interesting phenomena. Depending on the charges of LLPs, this leads to signatures of displaced vertices for neutral particles and disappearing tracks for charged particles.

Current collider searches for LLPs utilize a traditional detector located at the interaction point (IP).
In our previous work~\cite{Wang:2019xvx}, inspired by the proposed new experiments MATHUSLA~\cite{Chou:2016lxi,Curtin:2018mvb}, CODEX-b~\cite{Gligorov:2017nwh}, FASER~\cite{Feng:2017uoz}, AL3X~\cite{Gligorov:2017nwh} and ANUBIS~\cite{Bauer:2019vqk}, we have proposed the installation of FAr Detectors at the Electron Positron Collider (FADEPC), which are new detectors at a position far from the IP at generic high energy $e^-e^+$ colliders such as the Circular Electron Positron Collider (CEPC)~\cite{CEPCStudyGroup:2018rmc, CEPCStudyGroup:2018ghi, CEPCAcceleratorStudyGroup:2019myu}, the $e^-e^+$ running mode of the Future Circular Collider (FCC-ee)~\cite{FCC:2018byv, FCC:2018evy}, the International Linear Collider (ILC)~\cite{Behnke:2013xla, Baer:2013cma, Phinney:2007gp, Behnke:2013lya} and the Compact Linear Collider (CLIC)~\cite{Linssen:2012hp, Klamka:2021cjt}
\footnote{Similar idea was also proposed later in Ref.~\cite{Chrzaszcz:2020emg}}.
Such new detector is called ``far detector'' or abbreviated as ``FD" in this article. 
We develop eight different designs of such far detectors by varying the locations, volumes, and geometries.
We investigate their discovery potential for three physics scenarios: SM Higgs bosons are produced at a center-of-mass energy of $\sqrt{s} = 240$ GeV and decay into a pair of long-lived scalars $h \to XX$; $Z-$bosons are produced at $\sqrt{s} = 91.2$ GeV and decay into either a long-lived heavy neutral lepton and an active neutrino $Z \to N\nu$, or a pair of long-lived lightest neutralinos $Z \to \neutralino1 \neutralino1$ in the context of RPV-SUSY.
The limits on the model parameters are given for both near detectors and far detectors at the CEPC and FCC-ee.
We find that when searching for LLPs, such new experiments with far detectors at future lepton colliders can extend and complement the sensitivity reaches of experiments at future lepton colliders with the usual near detectors and the present and future experiments at the LHC.

In this work, we present a strategy to detect long-lived axion-like particles (ALPs) and explore the discovery potential of various far detectors at future lepton colliders.
The original axion was introduced as a pseudo Nambu-Goldstone boson (pNGB) of the Peccei-Quinn symmetry broken by the axial anomaly of QCD to solve the strong CP problem in the SM strong interaction~\cite{PhysRevLett_38_1440, Peccei:1977ur, PhysRevLett_40_223, PhysRevLett_40_279, Kim:1979if, Shifman:1979if, Zhitnitsky:1980tq, Dine:1981rt, DiLuzio:2020wdo}.
This idea was later extended to ALPs which are generic pNGBs arising naturally from the spontaneous breaking of a new U(1) global symmetry at some large energy scale $\Lambda$ in many beyond standard model theories.
Since their masses and couplings to SM particles are model dependent and can range over many orders of magnitude, 
ALPs have been investigated extensively in different regions of the parameter space spanned by the ALP masses and couplings.

The early studies on the ALP searches at colliders can be seen in Refs.~\cite{Mimasu:2014nea,Kleban:2005rj}.
Depending on their masses and couplings to SM particles, ALPs can have different production processes and decay modes at colliders. Therefore,  the search strategies for ALPs also vary greatly at different collider experiments.
In Ref.~\cite{Belle-II:2020jti}, the Belle II collaboration search for the signal process $e^- e^+ \to \gamma a, a \to \gamma \gamma$ and constrain the ALP-photon-photon coupling in the mass range $0.2 < m_a < 9.7$ GeV.
At the LHC, the ALPs can be produced from the Higgs boson decays $h \to a a / a Z$, the gluon-gluon fusion process $ g g \to a$, the associated boson processes $p p \to \gamma a / W a / Z a / h a$ or other vector boson fusion
processes $\gamma \gamma / \gamma Z / Z Z / W^+ W^-\to a$, etc.
The CMS and ATLAS collaborations have performed analyses for the signal process $h \to a a$, with various decay modes $a \to \gamma\gamma$~\cite{ATLAS:2015rsn}, $a \to \mu^+ \mu^-$~\cite{CMS:2012qms}, $a \to \ell^+ \ell^-, \ell = e, \mu$~\cite{ATLAS:2018coo,CMS:2020bni}, $a \to \tau^+ \tau^-$~\cite{CMS:2015twz}, $a \to \mu^+ \mu^-, \tau^+ \tau^-, b \bar{b}$~\cite{CMS:2017dmg}, and $a \to \mu^+ \mu^-, b \bar{b}$~\cite{ATLAS:2021hbr}.
For ALPs produced from the Higgs boson decay $h \to a Z$, ATLAS and CMS collaborations have searched the final states with $a \to \ell^+ \ell^-, \ell = e, \mu$~\cite{ATLAS:2018coo, CMS:2020bni}, and $a \to g g, q \bar{q}$~\cite{ATLAS:2020pcy}.
Furthermore, the CMS collaboration has searched the signal processes $ g g \to a \to Z Z, Z h$~\cite{CMS:2021xor}, and $p p \to W a, a \to W W$~\cite{CMS:2019ruu}, while the ATLAS collaboration has also analyzed the photon-photon fusion process $\gamma \gamma \to a \to \gamma \gamma$ in the Pb-Pb collision data and constrained the ALP-photon-photon coupling in the mass range $6 < m_a < 100$ GeV.

Recent phenomenology studies on ALPs at $e^- e^+$ colliders can be found in Refs.~\cite{deBlas:2018mhx, Frugiuele:2018coc, Inan:2020kif, Comelato:2020cyk, Yue:2021iiu, Cheung:2021mol, Sakurai:2021ipp}, and more studies can be found in the reviews~\cite{Bauer:2017ris, Dolan:2017osp, Bauer:2018uxu, Zhang:2021sio, dEnterria:2021ljz, Agrawal:2021dbo} and references therein.

Since the leading ALPs' couplings to SM particles scale as $1/\Lambda$, their lifetime can be long for large $\Lambda$ and small $m_a$.
Such long-lived ALPs can be prime targets for proposed experiments with new far detectors~\cite{Curtin:2018mvb, MATHUSLA:2018bqv, MATHUSLA:2020uve, FASER:2018eoc, FASER:2019aik, Feng:2018pew, Dreyer:2021aqd} and other new experimental approaches~\cite{Irastorza:2018dyq, Berlin:2018bsc, Abramowicz:2021zja, Bai:2021dgm}.

We note that to optimize the design and realize the construction of such FDs at future lepton colliders, it is important to investigate sensitivities of different FD designs to various signals with typical production and decay modes.
This paper focuses mainly on the physics potential of various FD designs in the context of long-lived ALPs. 
The FDs are considered to have different locations, volumes and geometries to estimate the effects of such factors on physics discovery limits. 
Since the detector designs are just tentative proposals and the technologies are still under development, the details of technology and cost of FDs are  beyond the scope of this paper.

The article is organized as follows. 
In Sec.~\ref{sec:theory}, we present the theoretical aspects and formulation of the signal.
In Sec.~\ref{sec:simAna}, we describe the data simulation and the analysis strategy.
In Sec.~\ref{sec:results}, we state the results of the average decay probability and the limits on the model parameters.
We summarize and conclude in Sec.~\ref{sec:results}.

\section{Theory Models}
\label{sec:theory}

The general ALP effective Lagrangian including interactions with the SM electroweak gauge bosons can be 
written as \cite{Georgi:1986df}
\begin{eqnarray}
{\cal L}_{\rm eff}
		&\supset & \, \frac12 \left( \partial_\mu a\right)\!\left( \partial^\mu a\right)
		- \frac{m_{a}^2}{2}\,a^2
		+ g^2\,C_{WW}\,\frac{a}{\Lambda}\,W_{\mu\nu}^A\,\tilde W^{\mu\nu,A} \nonumber \\
		&+& g^{\prime\,2}\,C_{BB}\,\frac{a}{\Lambda}\,B_{\mu\nu}\,\tilde B^{\mu\nu} \,,
\end{eqnarray}
where $W_{\mu\nu}^A$ and $B_{\mu\nu}$ denote the field strength tensors of the SU$(2)_L$ and U$(1)_Y$ gauge groups, and their  dual field strength tensors are defined as $\tilde X^{\mu\nu}=\frac12\epsilon^{\mu\nu\alpha\beta} X_{\alpha\beta}(X = W, B)$. The $g$ and $g'$ are coupling constants for each gauge group. 
The parameters $m_a$ and $\Lambda$ denote the ALP mass and the characteristic energy scale of the global symmetry breaking, which we assume to be independent parameters throughout this article.

After electroweak symmetry breaking, the effective Lagrangian including the interactions
of the ALP $a$ with $\gamma\gamma$, $\gamma Z$ and $ZZ$ is~\cite{Bauer:2018uxu}
 \begin{equation}\label{}
 	\begin{aligned}
 	{\cal L}_{\rm eff}
 	\supset&\, e^2\,C_{\gamma\gamma}\,\frac{a}{\Lambda}\,F_{\mu\nu}\,\tilde F^{\mu\nu}
 	+ \frac{2e^2}{s_w c_w}\,C_{\gamma Z}\,\frac{a}{\Lambda}\,F_{\mu\nu}\,\tilde Z^{\mu\nu}  \\
 	&+ \frac{e^2}{s_w^2 c_w^2}\,C_{ZZ}\,\frac{a}{\Lambda}\,Z_{\mu\nu}\,\tilde Z^{\mu\nu} \,.
 	\end{aligned}
 \end{equation}
 where $F_{\mu\nu}$ and $Z_{\mu\nu}$ are the field strength tensors of the electromagnetic field and $Z$ field, respectively. $s_w$ and $c_w$ are the sine and cosine of Weinberg angle, and $ e = g s_w$. The relevant Wilson coefficients are defined as~\cite{Bauer:2018uxu}
\begin{eqnarray}
 	C_{\gamma\gamma} &=& C_{WW}+C_{BB}, \notag \\
 	C_{\gamma Z} &=& c^2_w\,C_{WW}-s^2_w\,C_{BB}, \notag \\
 	C_{ZZ} &=& c^4_w\,C_{WW}+ s^4_w\,C_{BB}\, .
\end{eqnarray}

\begin{figure}[h]
	\centering
	\includegraphics[height=4cm, width=6cm]{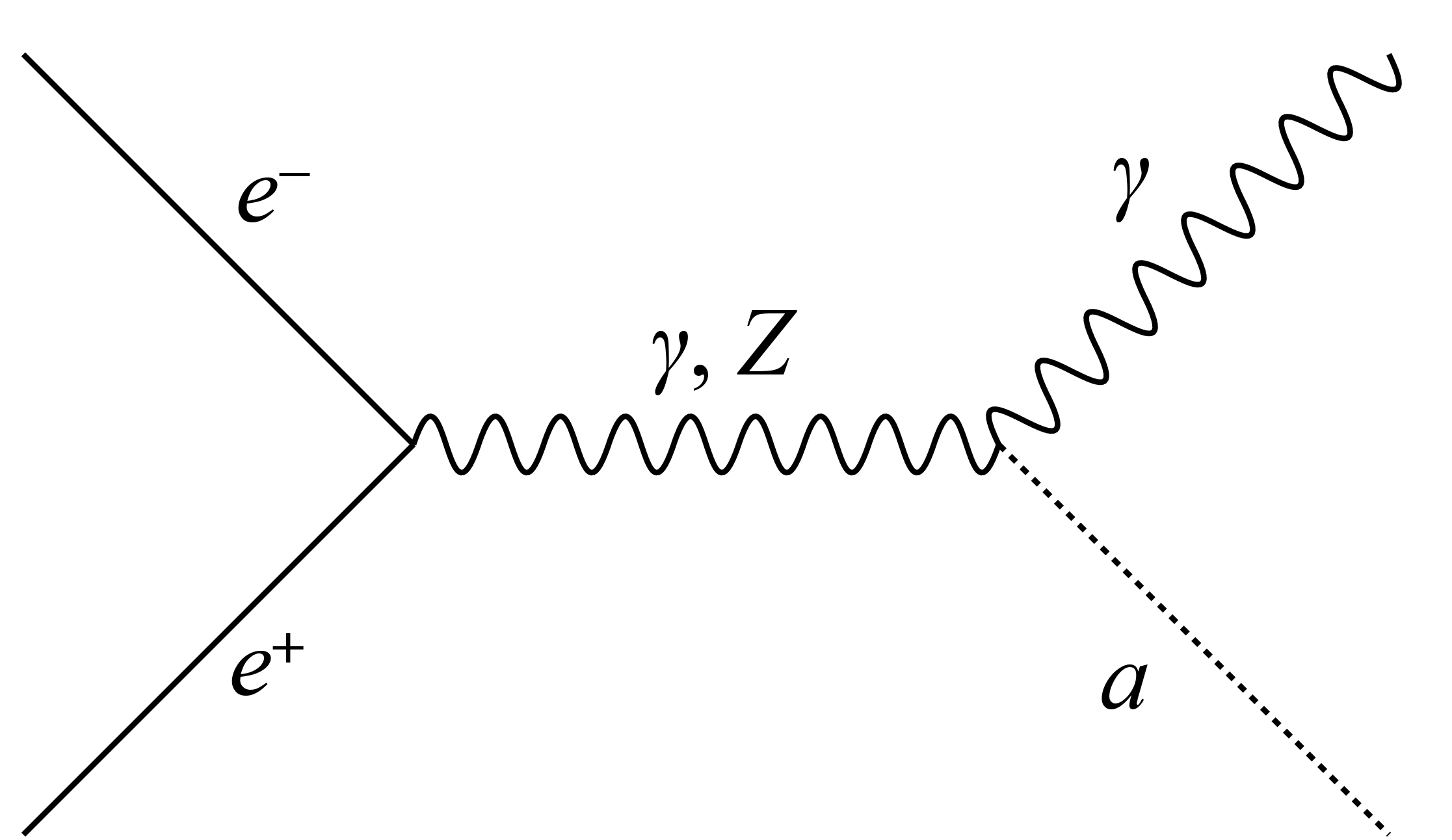}
	\caption{
	The production process of $e^- e^+ \to   \gamma \, a$ at electron-positron colliders.
	}
	\label{fig:eegga}
\end{figure}

Assuming ALPs interact with the electroweak gauge bosons~\footnote{The ALP with coupling to photons may play the role of the inflaton~\cite{Takahashi:2019qmh}.}, at $e^+e^-$ colliders, ALPs can be produced in association with one photon or $Z-$boson, or produced from the exotic decays of $Z-$bosons.
In this work, we consider the ALP production associated with one photon. The corresponding process is $e^-e^+ \rightarrow \gamma \,\, a$, where $a$ is an ALP, and
the production process is shown in Fig.~\ref{fig:eegga}.

The corresponding differential cross section is calculated in Ref.~ \cite{Bauer:2018uxu} as
\begin{align}
	\frac{d \sigma(e^- e^+\to\gamma a)}{d\Omega} =&
	\, 2 \pi \alpha \alpha^2(s) \frac{s^2}{\Lambda^2} \left( 1 - \frac{m_a^2}{s} \right)^3 \left(1 + \cos^2 \theta \right) \notag \\
	& \times \left( |V_\gamma(s)|^2 + |A_\gamma(s)|^2 \right),
\label{eqn:diffCrs}
\end{align}
where 
\begin{align}
	V_\gamma(s) &= \frac{C_{\gamma \gamma}}{s} + \frac{g_V}{2 c_w^2 s_w^2}\frac{C_{\gamma Z}}{s - m_Z^2+ i m_Z \Gamma_Z}\, , \notag\\ 
	A_\gamma(s) &= \frac{g_A}{2 c_w^2 s_w^2}\frac{C_{\gamma Z}}{s - m_Z^2+ i m_Z \Gamma_Z}\, , 
\end{align}
with $g_V = 2 s_w^2 - 1/2$ and $g_A=-1/2$.

When the ALP mass is below the $Z-$boson mass, it mainly decays to a pair of photons with the decay width~\cite{Bauer:2017ris, Bauer:2018uxu}
\begin{equation}
\Gamma(a\to\gamma\gamma)  = 4\pi\alpha^2 m_a^3 \, \left| \frac{C_{\gamma\gamma} }{\Lambda} \right|^2 \, .
\label{eqn:width}
\end{equation}

\section{Simulation and Analysis}
\label{sec:simAna}

Similar to our previous work~\cite{Wang:2019xvx}, we consider the CEPC and FCC-ee as the benchmark lepton colliders.
As a $Z-$factory with $\sqrt{s} =$ 91.2 GeV, the CEPC would run in 2 years with two IPs, corresponding to a total integrated luminosity of $\mathcal{L}_Z^{\text{CEPC}}= 16\,\, \iab$~\cite{CEPCStudyGroup:2018ghi}, while the FCC-ee is designed to run in 4 years with two IPs, corresponding to a total integrated luminosity of $\mathcal{L}_Z^{\text{FCC-ee}}= 150\,\, \iab$~\cite{Abada:2019zxq}.

We apply the ALP model file with the linear Lagrangian~\cite{Brivio:2017ije}
\footnote{
Ref.~\cite{Brivio:2017ije} uses different notation for the model parameters (i.e. $c_{\tilde{W}}$,  $c_{\tilde{B}}$, and  $f_a$). 
In terms of the symbols in Ref.~\cite{Brivio:2017ije}, $C_{\gamma\gamma} / \Lambda = - ( c_w^2 c_{\tilde{B}} + s_w^2 c_{\tilde{W}}  ) / (f_a\,e^2)$, $C_{\gamma Z} / \Lambda = c_w^2 s_w^2 ( c_{\tilde{B}} - c_{\tilde{W}} ) / (f_a e^2)$.
} 
in the Universal FeynRules Output (UFO) format~\cite{Degrande:2011ua} into the MadGraph5 program~\cite{Alwall:2014hca} to simulate the electron-positron collisions and generate the $e^-e^+ \to \gamma \,\, a$ events.  
The decays of ALPs are performed by PYTHIA8~\cite{Sjostrand:2006za, Sjostrand:2014zea}, and the data are output in the HEPMC~\cite{Dobbs:2001ck} format.
To maintain consistency throughout our study, the production cross sections calculated by MadGraph5 are used to estimate the number of signal events.
We vary the model parameters $m_a$, $C_{\gamma\gamma}$,  $C_{\gamma Z}$, and $\Lambda$ and calculate the production cross sections numerically using MadGraph. The numerical expression of production cross section is found to be
\begin{eqnarray}
\sigma(e^-&&e^+ \to \gamma\,a) 
\approx 16 \fb \times \left( \frac{\TeV}{\Lambda} \right)^2  \left( 1 - \frac{m_a^2}{s} \right)^3 \nonumber \\
&& \left(  \left| C_{\gamma\gamma} \right|^2 + 2680\left| C_{\gamma Z} \right|^2 - 0.082\left| C_{\gamma\gamma} C_{\gamma Z} \right|   \right)
\label{eqn:crs}
\end{eqnarray}
The dependences on the model parameters are compared with the theoretical expressions in Eqs.~\ref{eqn:diffCrs}, and are found to be consistent.

Similar to our previous work~\cite{Wang:2019xvx}, the total number of ALPs decaying in the fiducial volume can be calculated as
\begin{equation}
N_{\rm{ALP}}^{\rm{obs}} = N_{\rm{ALP}}^{\rm{prod}} \cdot \langle P[\rm{ALP\,in\,f.v.}]\rangle \cdot \rm{Br(ALP}\to\rm{visible} ).
\label{Eq:calSignal}
\end{equation} 
Here $N_{\rm{ALP}}^{\rm{prod}} = \sigma(e^- e^+ \rightarrow \gamma a)\times \mathcal{L}_Z$ is the total number of ALPs produced at future lepton colliders, where $\mathcal{L}_Z$ denotes the integrated luminosity.
$\langle P[\rm{ALP\,in\,f.v.}]\rangle$ stands for the average decay probability of the ALPs inside the detector's fiducial volume.
$\rm{Br(ALP}\to\rm{visible)}$ represents the branching ratio of the ALP decaying into visible final state.

The average decay probability is computed with the following procedure
\begin{eqnarray}
	\langle P[\text{ALP}\text{ in f.v.}]\rangle=\frac{1}{N^{\text{MC}}_{\text{ALP}}}\sum_{i=1}^{N^{\text{MC}}_
		{\text{ALP}}}P[(\text{ALP})_i\text{ in f.v.}]\,,
\end{eqnarray}
Here $N^{\text{MC}}_{\text{ALP}}$ denotes the total number of ALPs generated with the Monte Carlo simulation tool MadGraph5 and PYTHIA8, while $P[(\text{ALP})_i\text{ in f.v.}]$ is the individual decay probability of the i-th ALP and is determined by the detector's geometries and its position relative to the IP. 
Considering the ALP is produced at IP and decays with exponential law as it travels straightly, the individual decay probability can be estimated as
\begin{equation}
P[(\text{ALP})_i\text{ in f.v.}] = e^{\left( -{ D_i^\text{first} }/{ \lambda_i } \right)} - e^{\left( -{D_i^\text{last}}/{\lambda_i} \right)}\, ,
\label{eqn:decayProb}
\end{equation}
where $D_i^\text{first}$ and $D_i^\text{last}$ represent the distances relative to the IP when the ALP firstly enters and lastly leaves the detector, respectively, 
and $\lambda_i$ is the decay length of the ALP in the laboratory frame.
Based on the kinematic information of each ALP provided by PYTHIA8, we derive the kinematic variables as follows:
$\beta_i = p_i / E_i$, $\gamma_i = E_i / m_a$, $\lambda_i = \beta_i \gamma_i c \tau$, 
where $p_i$ and $E_i$ are the momentum and energy of the i-th ALP, respectively.
$m_a$ is the ALP mass, and
$c\tau$ is its proper decay length, where $\tau$ is its lifetime in the rest frame and $c$ is the speed of light.

In this work, to exploit the high luminosities, we consider ALPs are produced through the $e^-e^+ \to \gamma\, a$ process at the $Z$-factory mode with $\sqrt{s} = 91.2$ GeV.
Based on energy-momentum conservation, the momentum $p_i = (s-m_a^2)/(2\sqrt{s})$, 
so that $\beta_i \gamma_i=p_i/m_a= (s-m_a^2)/(2 m_a \sqrt{s})$.
For the ALP masses considered in this article, the ALP decays mainly to a photon pair and we assume the total decay width $\Gamma_a = \Gamma (a\to \gamma\gamma)$.
Combined with Eq.~(\ref{eqn:width}), 
the decay length is calculated as
\begin{equation}
\lambda_i \approx 15\,\mathrm{m} 
 \left( \frac{s-m_a^2}{m_a \sqrt{s}} \right) 
  \left( \frac{\GeV}{m_a} \right)^3
\left( \frac{\Lambda}{\TeV} \right)^2 \left( \frac{10^{-4}}{C_{\gamma\gamma}} \right)^2 \, ,
\label{eqn:lambda}
\end{equation}
and depends on the parameters $m_a$ and $C_{\gamma\gamma}/\Lambda$.

\section{Results}
\label{sec:results}

\subsection{Average Decay Probability}
\label{sec:ADP}

A far detector, different from the traditional near detector located at the IP, is one additional detector installed at a location far from the IP. 
If LLPs have relatively long decay lengths, they can have high probabilities of decaying inside the far detector.
Thus, the FD can enhance the potential sensitivity reaches. 
As mentioned in Sec.~\ref{sec:simAna}, the average decay probability of the LLPs inside the detector's fiducial volume is determined by the detector's geometries and its position relative to the IP.
Therefore, it is important to investigate the discovery potential of different FD designs.
In our previous work~\cite{Wang:2019xvx}, we have developed eight different designs of far detectors by varying the locations, volumes, and geometries, which we label as ``FD1-FD8''.
Please refer to this reference for details of different FD designs.
We note that due to its small geometry size, FD1 can be placed inside the experiment hall if the hall is big enough or it can be placed in a cavern or shaft near the experiment hall.
Other designs have big volumes and can be placed on the ground above the IP.

\begin{figure}[h]
	\centering
	\includegraphics[height=5cm, width=8cm]{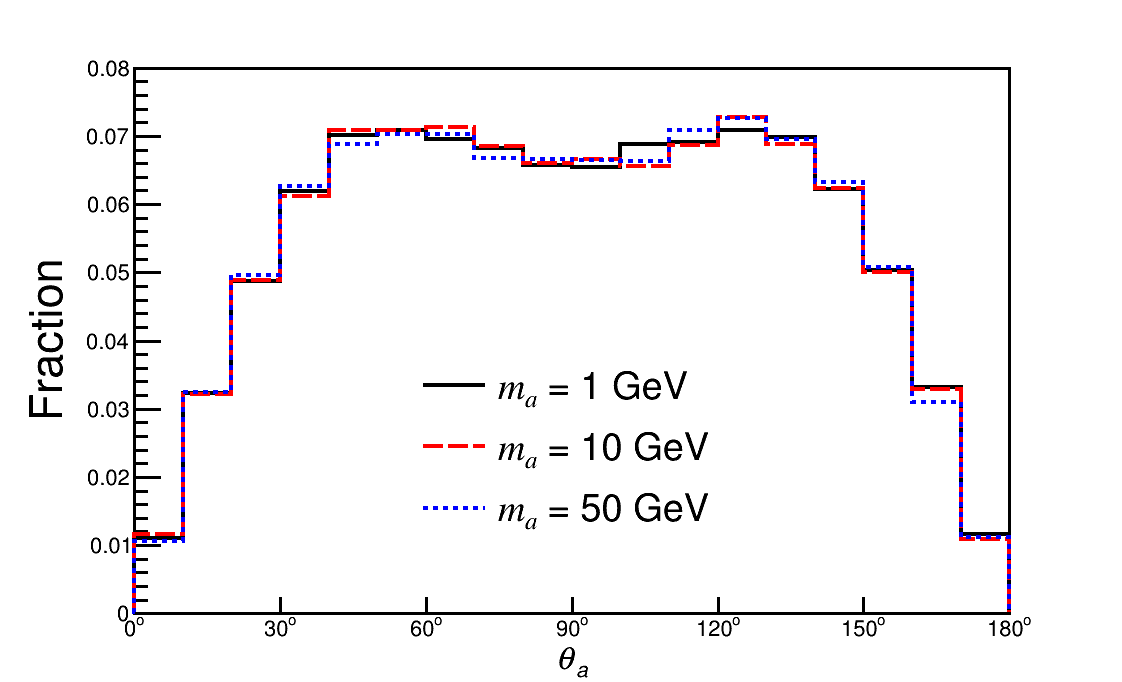}
	\caption{
	The angle $\theta$ distributions of ALPs produced from $e^-e^+ \to \gamma\, a$ process at future $e^-e^+$ colliders with $\sqrt{s} = 91.2$ GeV for three different ALP masses of 1, 10, 50 GeV.
	}
	\label{fig:theta}
\end{figure}

Fig.~\ref{fig:theta} shows the distributions of polar angle $\theta$ for ALPs produced from $e^-e^+ \to \gamma\, a$ process at future $e^-e^+$ colliders with $\sqrt{s} = 91.2$ GeV for three different ALP masses of 1, 10, 50 GeV.
Here, the $z$–axis in the laboratory frame is along the incoming electron and positron beams and the ``$+z$” is defined as the electron beam’s forward direction. The polar angle $\theta$ is defined as usual, taking the positive $z$–axis as its zero value. 
One can see that the angle $\theta$ distribution has two peaks around $90^\circ \pm 40^\circ$, and most ALPs travel nearly transversely.
This is consistent with the expression of differential cross section shown in Eq.~(\ref{eqn:diffCrs}). 
Besides, the angle $\theta$ distribution is insensitive to the ALP masses.

\begin{figure}[h]
\centering
\includegraphics[height=6cm, width=9cm]{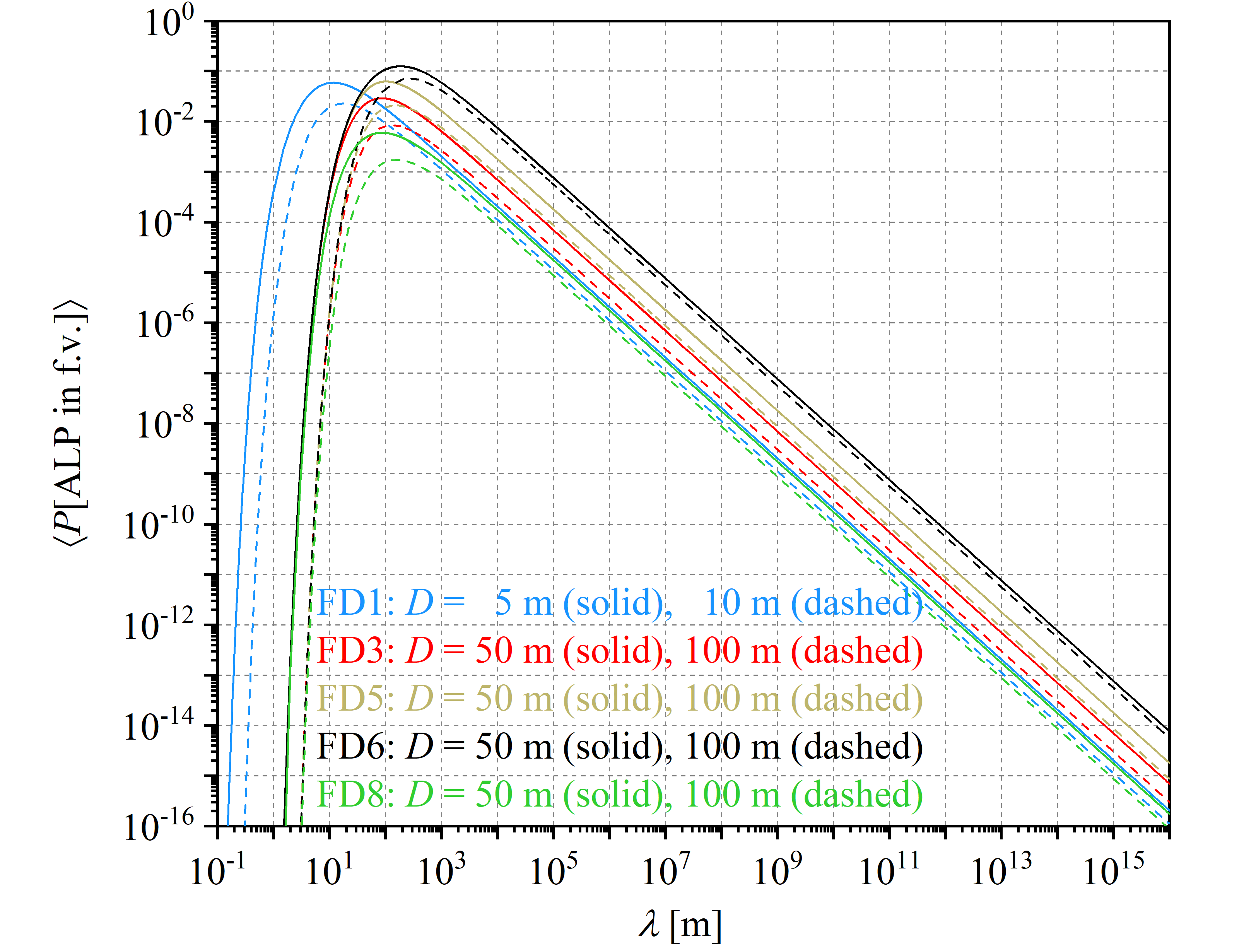}
\caption{
The average decay probability of various far detectors FDi ( i = 1, 3, 5, 6, 8) with different choices of $D$ as a function of ALP decay length $\lambda$.
}
\label{fig:ADPvsLambda}
\end{figure}

In Fig.~\ref{fig:ADPvsLambda}, we present the curves showing the relationship between the decay length $\lambda$ and the average decay probability $\langle P[\text{ALP}\text{ in f.v.}]\rangle$ of ALPs for far detectors FD1, FD3, FD5, FD6 and FD8 with two choices of $D$, where $D$ stands for the radial/transverse distance between the IP and the far detector.
Comparing the solid with the dashed curves, one can observe that 
the smaller $D$ can give higher probability for every FD. 
This is mainly because the closer distance to the IP is beneficial for receiving more ALPs in FD's direction. 

\begin{table}[h]
\begin{tabular}{cccc}
\hline
\hline
                     & $D$ [m]  & $\lambda$ [m] & $\langle P[\text{ALP}\text{ in f.v.}]\rangle$ \\
\hline                      
\multirow{2}{*}{FD1} & 5   &   12     &   $5.9 \times 10^{-2}$  \\
                     & 10  &   18     &  $2.3 \times 10^{-2}$   \\
\hline                     
\multirow{2}{*}{FD3} & 50  &   87     &  $2.9 \times 10^{-2}$   \\
                     & 100 &   132     &   $8.4 \times 10^{-3}$  \\
\hline                      
\multirow{2}{*}{FD5} & 50  &  103      &  $6.2 \times 10^{-2}$   \\
                     & 100 &  151      &  $2.1 \times 10^{-2}$   \\
\hline                      
\multirow{2}{*}{FD6} & 50  &   185     &  $1.2 \times 10^{-1}$    \\
                     & 100 &   262     & $7.2 \times 10^{-2}$    \\
\hline                      
\multirow{2}{*}{FD8} & 50  &   87     &  $6.0 \times 10^{-3}$   \\
                     & 100 &    158    & $1.7 \times 10^{-3}$   \\
\hline
\hline                     
\end{tabular}
\caption{The peak coordinates for curves in Fig.~\ref{fig:ADPvsLambda}.}
\label{tab:maxP}
\end{table}

We find that the performance of FD4 is almost identical to FD3, while the performance of FD7 is slightly weaker than FD8.
The limit for FD2 is between FD8 and FD3 when $\lambda \gtrsim 100$ m, while FD2's limit is weakest when $\lambda \lesssim 100$ m.
FD1 has the biggest probability when $\lambda \lesssim 40$ m, while FD6 has the biggest probability when $\lambda \gtrsim 40$ m.
To show the maximal values of the average decay probability explicitly for all FDs, we list the peak coordinates for all curves in Table~\ref{tab:maxP}.
The decay lengths $\lambda$ corresponding to peak values are slightly higher than $D$.
Since FD6 has much larger volume and thus more space for accepting the decaying ALPs, its peak probability can reach $1.2 \times 10^{-1}$ with $D$ = 50 m .

It is worth noting that the behaviors of the curves in Fig.~\ref{fig:ADPvsLambda} can be understood qualitatively based on the properties of Eq.~(\ref{eqn:decayProb}).
To check the tendency of the curve, one can take the first derivative of Eq.~(\ref{eqn:decayProb}) and obtain 
\begin{equation}
\frac{\text{d\,log} P}{\text{d\,log} \lambda_i } = \frac{1}{{\lambda_i}}\left[ D_i^\text{first} -\frac{D_i^\text{last}-D_i^\text{first}  }{e^{(D_i^\text{last}-D_i^\text{first})/\lambda_i} -1 } \right] \, .
\label{eqn:derlogdecayProb}
\end{equation}
The peak position lies at $\lambda \sim \lambda_p = (D_i^\text{last}-D_i^\text{first}) / \text{ln\,} (D_i^\text{last}/D_i^\text{first})$, which corresponds to $\text{d\,log} P / \text{d\,log} \lambda_i = 0$.
When $\lambda$ is smaller (bigger) than $\lambda_p$, the value of Eq.~(\ref{eqn:derlogdecayProb}) is positive (negative), which corresponds to the increasing (decreasing) of the curves.
When $\lambda \gg \lambda_p$, $\text{d\,log} P / \text{d\,log} \lambda_i$ becomes one fixed value of about -1, which explains the linear behaviour of the curves on the right side of peak position.
When $\lambda \ll \lambda_p$, $\text{d\,log} P / \text{d\,log} \lambda_i$ becomes much larger as $\lambda$ decreases.
Therefore, compared with the steadily descending tendency of the curves on the right side of the peak positions, the curves on the left side of the peak positions descend much faster as $\lambda$ decreases.
As shown later, the tendencies of the average decay probability $\langle P[\text{ALP}\text{ in f.v.}]\rangle$ vs. $\lambda$ curves affect the behaviours of the limit boundaries greatly.

\subsection{Sensitivities on Model Parameters}
\label{sec:colliderSen}

Similar to our previous work~\cite{Wang:2019xvx}, since the aim of this study is to estimate the discovery potential of different FD designs, to simplify the analyses, we assume that the final state photons from ALP decays are detectable in the FDs, and $\text{Br}(\text{ALP} \to \text{visible} ) = 1$.
We also assume that backgrounds can be reduced to negligible levels for all FDs,
and present the sensitivity results in terms of 3-signal-event contour curves which correspond to 95\% C.L. limits with zero background events.
We note that the sensitivity limits would be reduced to some extent according to future realistic detector efficiency and background studies.
Because the detector designs are just tentative proposals and the technologies are still under development, we leave the more realistic results for future studies.
To probe the dependence on all model parameters $m_a$, $C_{\gamma\gamma}$, $C_{\gamma Z}$, and $\Lambda$, we show our results in the following three different cases.
For each plot, the parameter regions inside the contour curves have more than three signal events and are discoverable with 95\% CL with the background free assumption. 

We note that this paper considers not only the traditional case where the ALP couples to two photons via $C_{\gamma\gamma}$ only, i.e. the $C_{\gamma Z} =0$ case, but also another two cases:  $C_{\gamma Z} = C_{\gamma\gamma}$, or $C_{\gamma Z}$ and $C_{\gamma\gamma}$ can freely change.
Since the signal production cross section depends strongly on $C_{\gamma Z}$, the effect of this model parameter is very important for this study.
Therefore, the results of the additional two cases are also significant for this signal process.

\subsubsection{$C_{\gamma Z} =0$}
\label{sec:cgz0}

\begin{figure}[h]
	\centering
	\includegraphics[height=6cm, width=9cm]{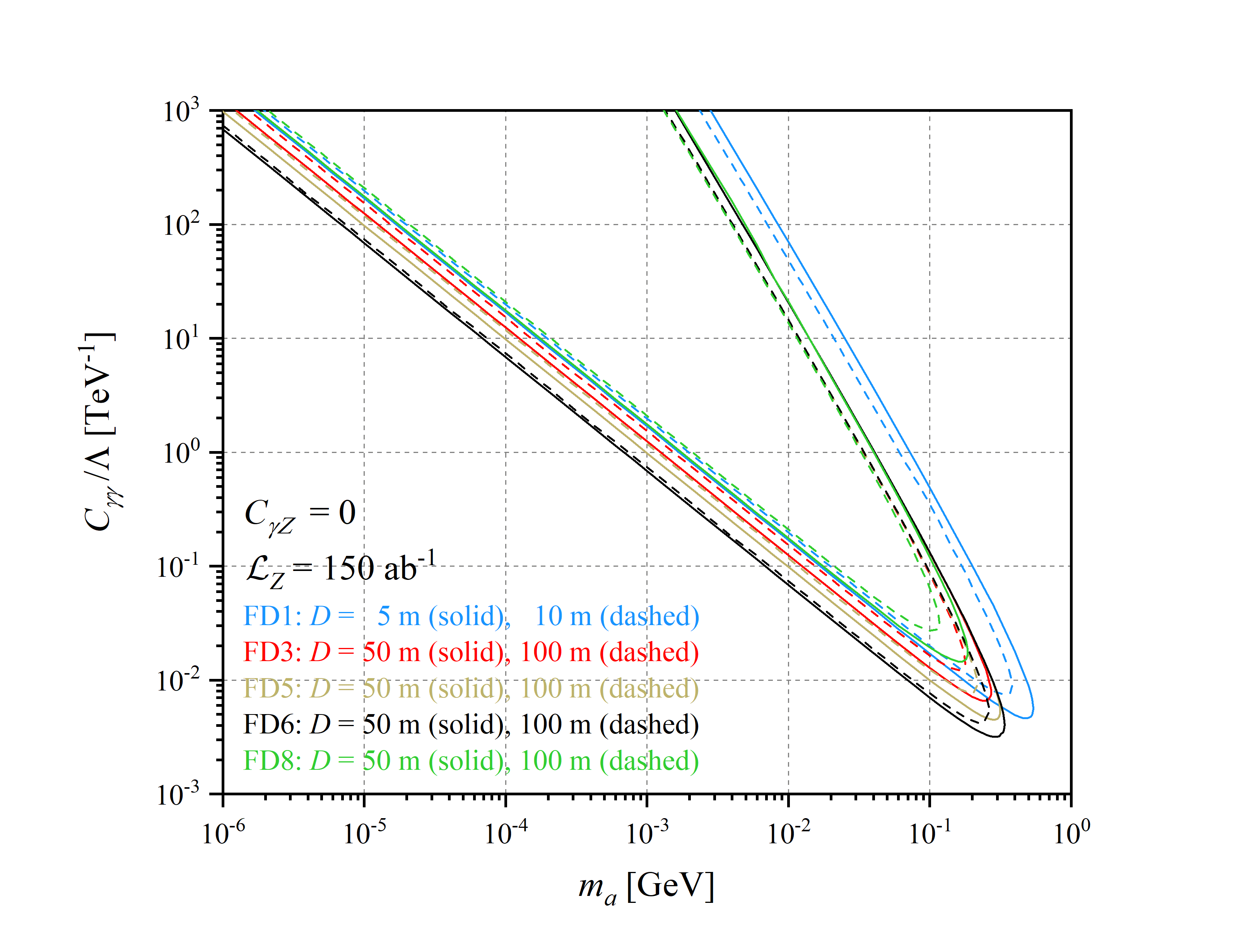}
	\includegraphics[height=6cm, width=9cm]{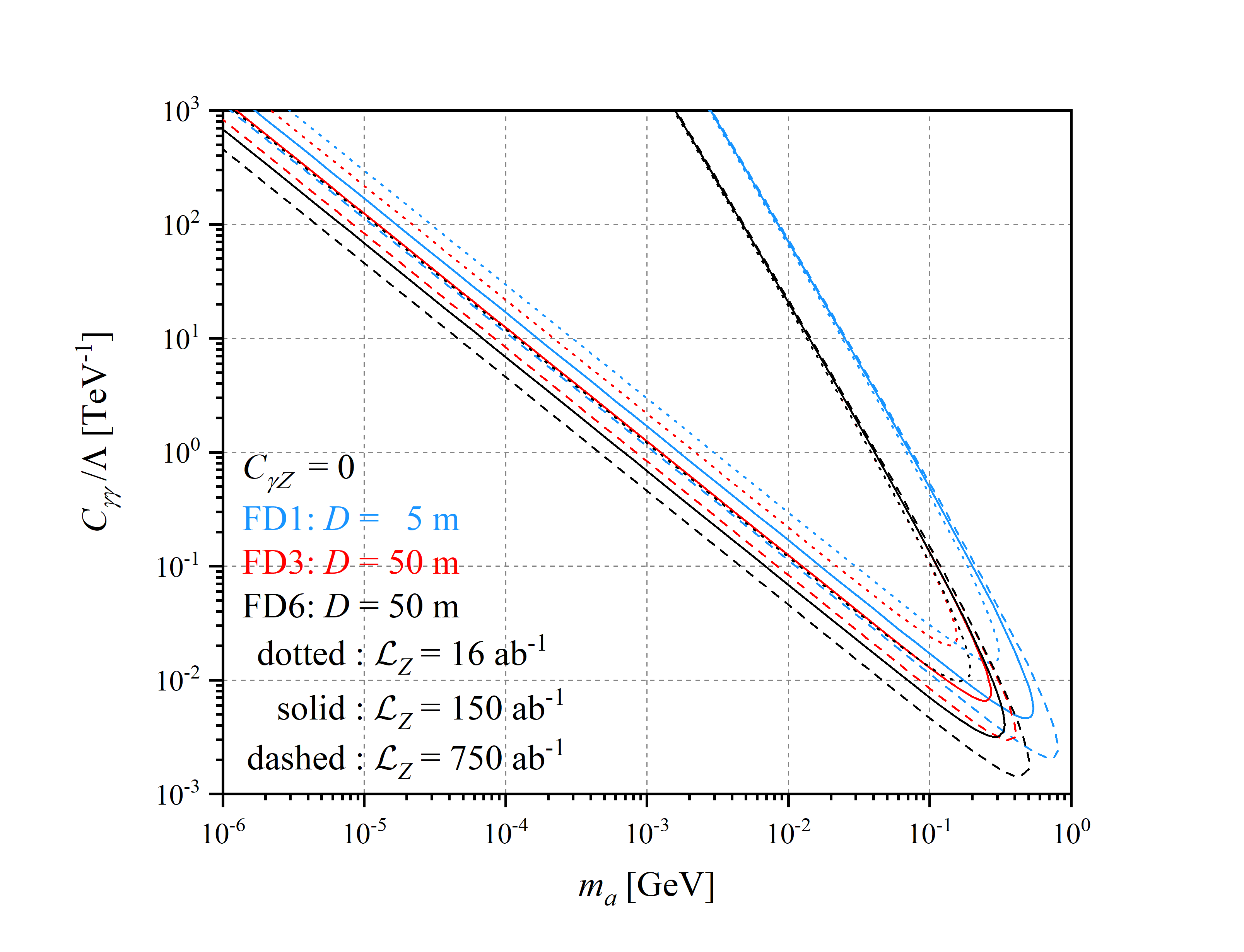}
	\caption{Upper: Sensitivity reaches of representative far detectors with different $D$ options and integrated luminosity of $\mathcal{L}_Z$ = 150 ab$^{-1}$ in the $C_{\gamma\gamma}/\Lambda$ vs $m_a$ plane. Lower: Sensitivity reaches of far detectors with different integrated luminosities $\mathcal{L}_{Z}$.}
	\label{fig:limitsCgz0}
\end{figure}

In this case, the model parameter $C_{\gamma Z}$ is assumed to be 0. 
The limits are presented in Fig.~\ref{fig:limitsCgz0} in the $C_{\gamma\gamma}/\Lambda$ vs. $m_a$ plane.

The upper plot shows the sensitivity reaches of the representative far detectors FD1, FD3, FD5, FD6, and FD8 for both options of $D$ with the integrated luminosity of $\mathcal{L}_Z$ = 150 ab$^{-1}$.
Comparing the solid with the dashed curves, one observes that the designs with smaller $D$ can cover wider region in the parameter space. 
This is because, as shown in Fig.~\ref{fig:ADPvsLambda} the designs with smaller $D$ have larger average decay probabilities.

One can also observe that FD1 has the largest mass reach to $m_a  = 0.54 $ GeV with $C_{\gamma\gamma}/\Lambda = 5.5\times10^{-3}~\iTeV$. 
The mass reaches of FD3 and FD6 can be $\sim$ 0.2\,–\,0.4 GeV with $C_{\gamma\gamma}/\Lambda$ = 7.0$\times10^{-3}$ and 4.0$\times10^{-3}~\iTeV$, respectively. 
The sensitivity of FD5 is between those of FD3 and FD6. 
Compared with other detectors, FD6 can reach smaller $C_{\gamma\gamma} / \Lambda$ for almost the entire mass range, while FD1 can probe the regions with higher $C_{\gamma\gamma} / \Lambda$, and FD8 has the weakest discovery potential.
 
In the lower plot, we compare the performances of FD1, FD3, and FD6 with various integrated luminosities of $\mathcal{L}_Z = 16, 150$ and 750 $\iab$ which correspond to $\mathcal{L}_Z^{\text{CEPC}}$, $\mathcal{L}_Z^{\text{FCC-ee}}$ and $5 \mathcal{L}_Z^{\text{FCC-ee}}$, respectively.
Here we choose the value of $5 \mathcal{L}_Z^{\text{FCC-ee}}$ to demonstrate the change in the limits with increasing luminosities.
The FD1's limits on $C_{\gamma\gamma}/\Lambda$ can reach as low as $1.4\times10^{-2}, 4.6\times10^{-3}, 2.1\times10^{-3}~\iTeV $ for 16, 150, 750 $\iab$ luminosities, respectively, while the lowest limits on $C_{\gamma\gamma}/\Lambda$ are $2.0\times10^{-2}, 6.6\times10^{-3}, 3.0\times10^{-3}~\iTeV$ and $9.7\times10^{-3}, 3.2\times10^{-3}, 1.4\times10^{-3}~\iTeV$ for FD3 and FD6, respectively.

It is obvious that large luminosity is helpful to extend the lower side of the limit boundary of the parameter space, while the enhancement is not substantial for the upper side.
This is because the signal rate is proportional to the product of integrated luminosity, production cross section and average decay probability, i.e. $N_{\rm{ALP}}^{\rm{obs}}  \propto \mathcal{L}_Z \times  \sigma(e^- e^+ \rightarrow \gamma a) \times \langle P[\rm{ALP\,in\,f.v.}]\rangle$. As the luminosity increases by a factor of ten, for example, the product of $\sigma(e^- e^+ \rightarrow \gamma a) \times \langle P[\rm{ALP\,in\,f.v.}]\rangle$ needs to be reduced by the same factor. 
The lower side of the limit boundary has small $C_{\gamma \gamma} / \Lambda$ and thus big $\lambda$ values corresponding to the rightmost curve in Fig.~\ref{fig:ADPvsLambda}. As $C_{\gamma \gamma} / \Lambda$ decreases, the $\lambda$ value increases and the average decay probability $ \langle P[\rm{ALP\,in\,f.v.}]\rangle$ descends slowly, which combining with the deceasing of the production cross section $\sigma(e^- e^+ \rightarrow \gamma a)$ renders the product reduced by a factor of ten.
By comparison, the upper side of the limit boundary has big $C_{\gamma \gamma} / \Lambda$ and thus small $\lambda$ values corresponding to the leftmost curve in Fig.~\ref{fig:ADPvsLambda}. When $C_{\gamma \gamma} / \Lambda$ increases, the $\lambda$ value decreases and the average decay probability $ \langle P[\rm{ALP\,in\,f.v.}]\rangle$ descends rapidly, which solely can render the product reduced by a factor of ten.
In summary, the luminosity's different effects on the lower and upper sides of the limit boundary are mainly because the rates of change between $ \langle P[\rm{ALP\,in\,f.v.}]\rangle$ and $\lambda$ are different for big and small  $\lambda$ cases. The same reason can also explain the luminosity effects in the lower plots of Figs.~\ref{fig:limitsCeq} and~\ref{fig:limitsM1}.

To compare with other studies,
the experimental research~\cite{Belle-II:2020jti} and phenomenological works~\cite{Jaeckel:2015jla, Dobrich:2015jyk,Aloni:2019ruo} also search for ALPs and probe the ALP's coupling to photons. 
The Belle II experiment~\cite{Belle-II:2020jti} searches for short-lived ALPs via the signal process $e^- e^+ \to \gamma a, a \to \gamma \gamma$, and considers 
the background processes $e^-e^+ \to \gamma \gamma \gamma$, and $P\gamma$, where $P$ is a SM pseudo-scalar meson. 
The 95\% confidence level upper limits on $C_{\gamma \gamma} / \Lambda$ is set at the level of 1 $\iTeV$ in the mass range $0.2 < m_a < 9.7$ GeV.
Ref.~\cite{Jaeckel:2015jla} considers the same signal process $e^- e^+ \to a \gamma, a \to \gamma \gamma$ at the LEP, and also assumes that ALPs decay promptly. 
The upper limit on $C_{\gamma \gamma} / \Lambda$ is set at around 10 $\iTeV$ for $0.1 \lesssim m_a \lesssim 10$ GeV, and around 1 $\iTeV$ for $10 \lesssim m_a \lesssim 90$ GeV. 
Ref.~\cite{Dobrich:2015jyk} studies the Primakoff production of long-lived ALPs in proton fixed target experiments.
Assuming no background
and $3.9 \times 10^{17}$ protons on target, the study predicts that the NA62 experiment would probe ALPs with mass below $\sim 200$ MeV, and the lowest $C_{\gamma \gamma} / \Lambda$ is $\sim 8.3 \times 10^{-3}~ \iTeV$.
The study also predicts that with $2 \times 10^{20}$ protons on target and negligible background, the proposed facility SHiP could extend the mass reach to around 1 GeV, and the lowest $C_{\gamma \gamma} / \Lambda$ is $\sim 5.6 \times 10^{-4}~ \iTeV$.
Ref.~\cite{Aloni:2019ruo} explores the sensitivity of photon-beam experiments to ALPs with QCD-scale masses.
This study considers that a photon beam is incident on a nuclear target, and ALPs are produced via the coherent Primakoff process.
The results forecast that the Pb data from the first PRIMEX run would probe ALPs in the mass range $0.03 \lesssim ma \lesssim 0.3$ GeV, and the lowest $C_{\gamma \gamma} / \Lambda$ is $\sim 0.56~ \iTeV$.
Because these studies have different assumptions, in order to avoid misleading, here we don't overlay other limits in Fig.~\ref{fig:limitsCgz0}.

\subsubsection{$C_{\gamma Z} = C_{\gamma \gamma}$}
\label{sec:cgzEqcgg}

\begin{figure}[h]
\centering
\includegraphics[height=6cm, width=9cm]{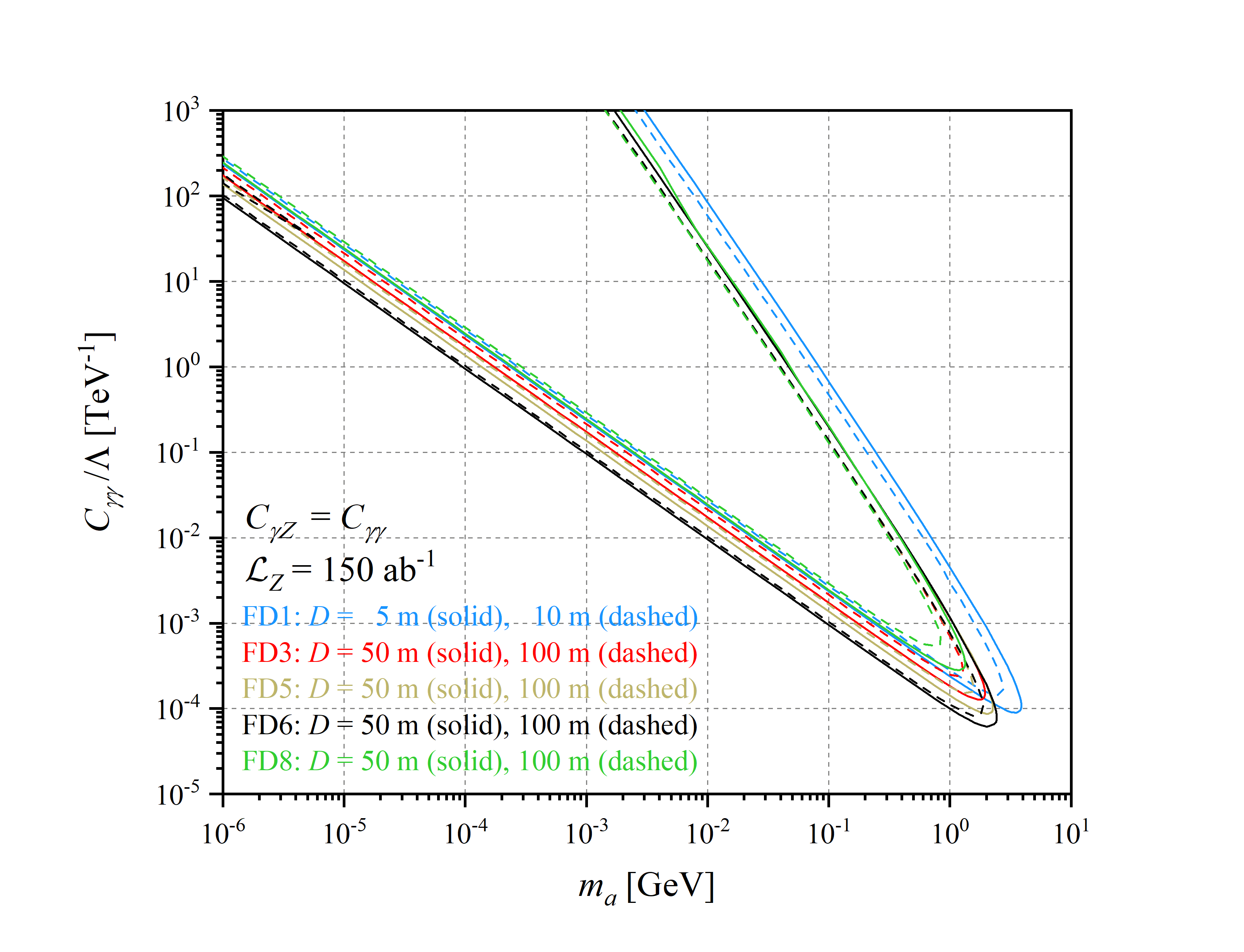}
\includegraphics[height=6cm, width=9cm]{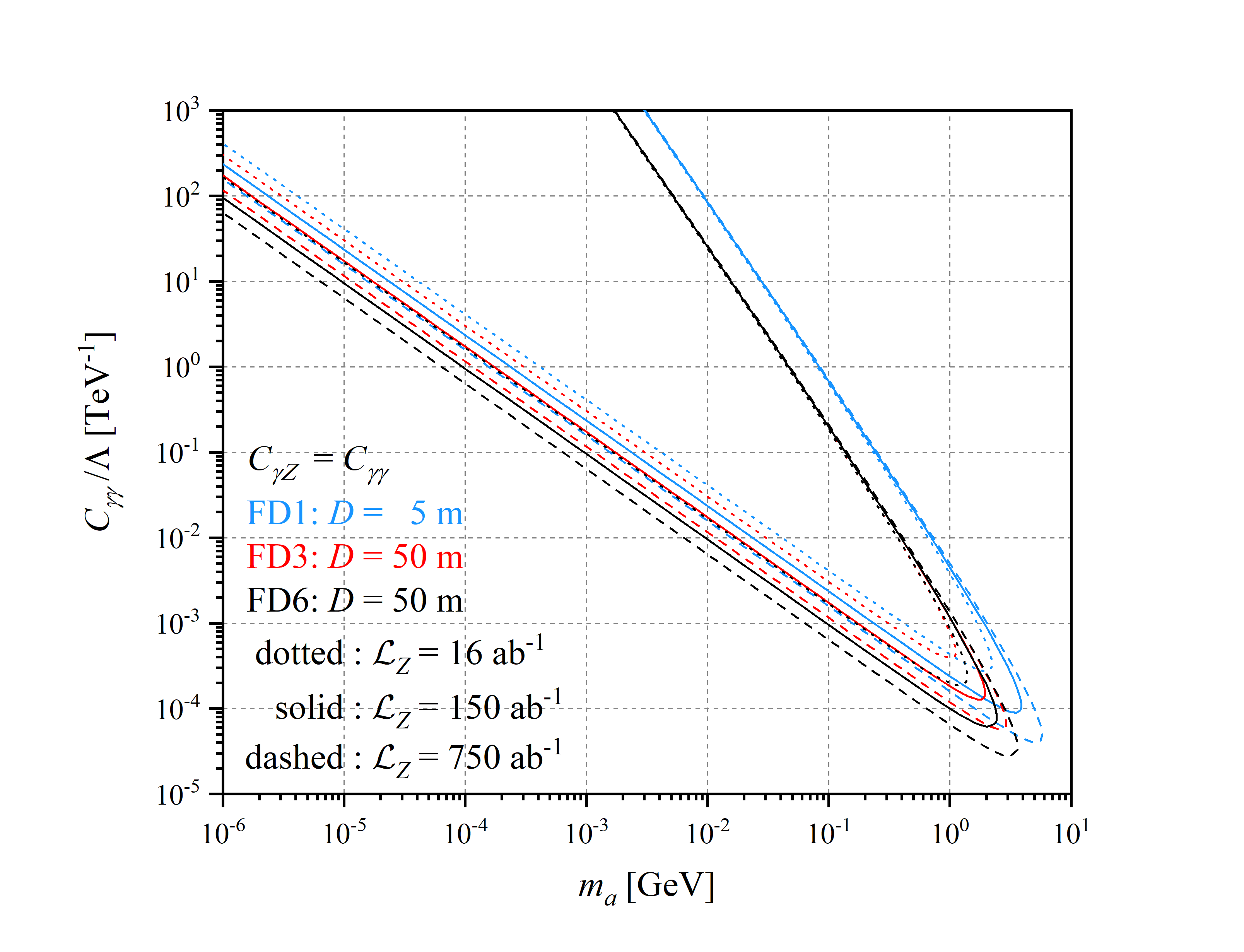}
\caption{The similar plots as Fig.\ref{fig:limitsCgz0} but for $C_{\gamma Z}$ = $C_{\gamma\gamma}$.}
\label{fig:limitsCeq}
\end{figure}

In this case, the model parameter $C_{\gamma Z}$ is assumed to be equal to $C_{\gamma\gamma}$. 
The limits are presented in Fig.~\ref{fig:limitsCeq} in the $C_{\gamma\gamma}/\Lambda$ vs. $m_a$ plane.
Similar to Fig.~\ref{fig:limitsCgz0}, the upper plot shows the sensitivity reaches of the representative far detectors with both options of $D$ and the integrated luminosity of $\mathcal{L}_Z$ = 150 ab$^{-1}$, 
while the lower plot shows the limits with various integrated luminosities of $\mathcal{L}_Z = 16, 150$ and 750 $\iab$.

Comparing the limits between Fig.~\ref{fig:limitsCeq} and Fig.~\ref{fig:limitsCgz0},  we find that the shapes of the discoverable parameter region are similar, but the coverages of the curves in Fig.~\ref{fig:limitsCeq} are larger. 
This is because, as shown in Eq.~(\ref{eqn:crs}) the model parameter $C_{\gamma Z}$ can greatly enhance the production cross sections.
In the upper plot, FD1 has the largest mass reach to $m_a =$ 4 GeV with $C_{\gamma\gamma}/\Lambda = 1.2\times10^{-4}~\iTeV$. 
The mass reaches of FD3 and FD6 can be $\sim$ 2\,–\,2.4 GeV with $C_{\gamma\gamma}/\Lambda$ = 1.5$\times10^{-4}$ and 8.0$\times10^{-5}~\iTeV$, respectively. 
In the lower plot, the FD1's limits on $C_{\gamma\gamma}/\Lambda$ can reach as low as $2.7\times10^{-4}, 9.0\times10^{-5}, 4.0\times10^{-5}~\iTeV $ for 16, 150, 750 $\iab$ luminosities, respectively, while the lowest limits on $C_{\gamma\gamma}/\Lambda$ are $4.0\times10^{-4}, 1.3\times10^{-4}, 5.7\times10^{-5}~\iTeV$ and $2.0\times10^{-4}, 6.2\times10^{-5}, 3.0\times10^{-5}~\iTeV$ for FD3 and FD6, respectively.
Therefore, the lowest reach on $C_{\gamma\gamma}/\Lambda$ in Fig.~\ref{fig:limitsCgz0} is around $10^{-3} - 10^{-2}~ \iTeV$, while it is around $10^{-5} - 10^{-4}~ \iTeV$ in Fig.~\ref{fig:limitsCeq}.  

The limits of FD8 are weaker than other detectors in Fig.~\ref{fig:limitsCgz0} and Fig.~\ref{fig:limitsCeq}. 
Comparing the designs of FD8 and FD3, they have the same volume and height of 20 m.
The main difference is the bottom surface: FD8 is 2000 m $\times$ 20 m while FD3 is 200 m $\times$ 200 m. The performance of FD3 is better than FD8, which means that increasing the length in the beam direction (from 200 m to 2000 m) can not increase its discovery potential for the ALP signal.

\subsubsection{Free $C_{\gamma Z}, C_{\gamma \gamma}$ Parameters}
\label{sec:cgzcgg}

In this case, both the model parameter $C_{\gamma Z}$ and $C_{\gamma\gamma}$ are assumed to vary freely. 

\begin{figure}[h]
\centering
\includegraphics[height=6cm, width=9cm]{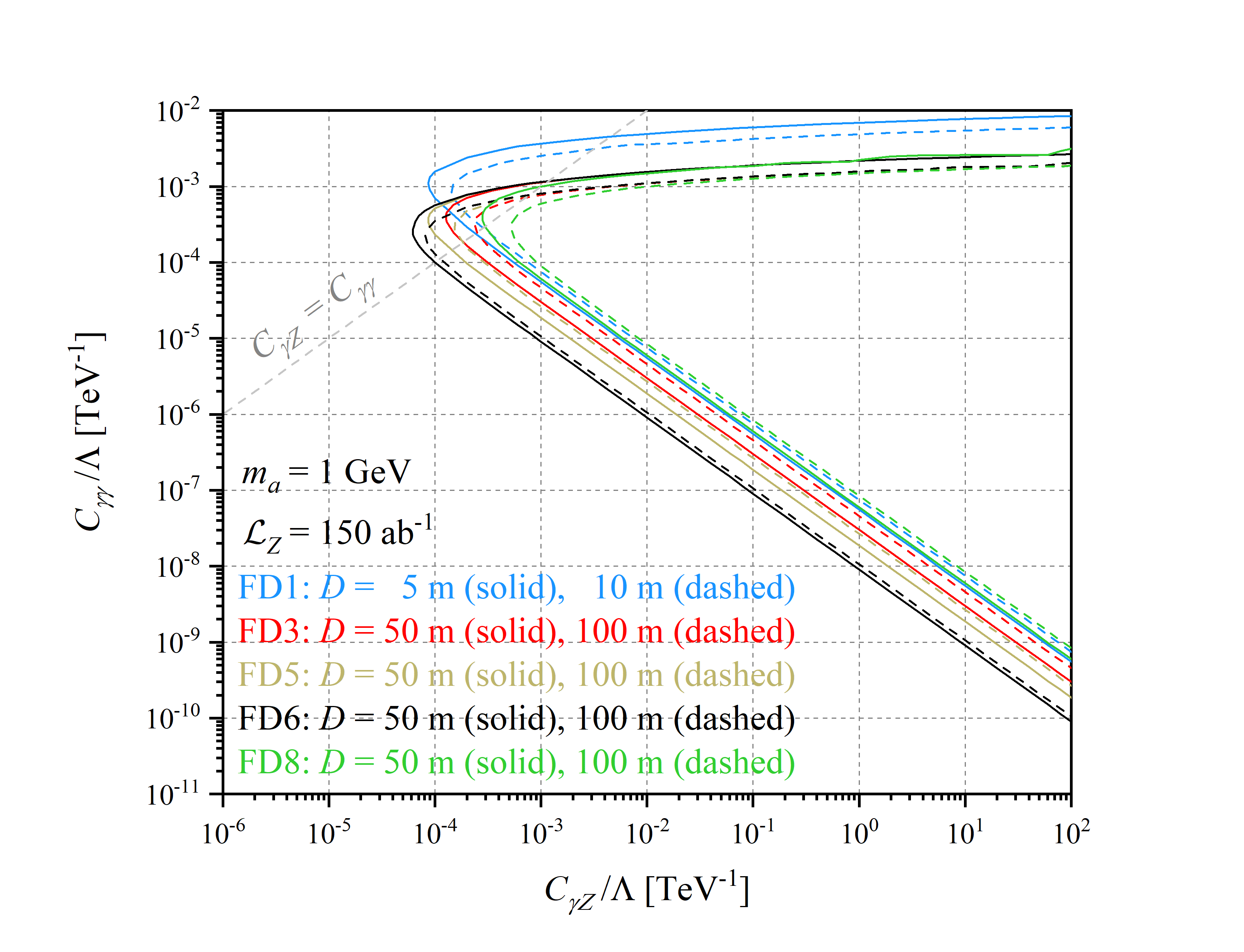}
\includegraphics[height=6cm, width=9cm]{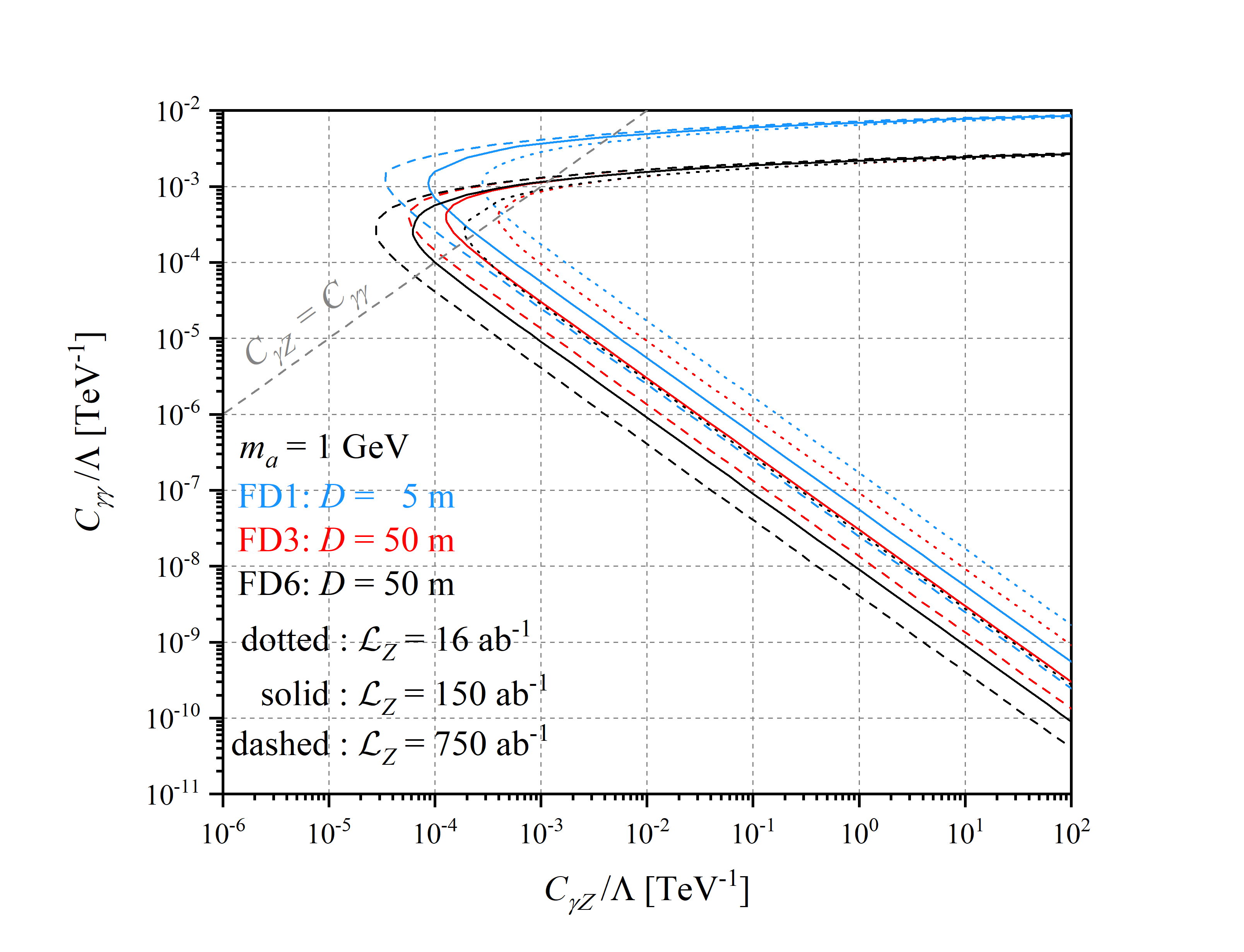}
\caption{
Upper: Sensitivity reaches of representative far detectors with different $D$ options and integrated luminosity of $\mathcal{L}_Z$ = 150 ab$^{-1}$ in the $C_{\gamma\gamma}/\Lambda$ vs $C_{\gamma Z}/\Lambda$ plane when $m_a = 1$ GeV. 
Lower: Sensitivity reaches of far detectors with different integrated luminosities $\mathcal{L}_{Z}$.
}
\label{fig:limitsM1}
\end{figure}

\begin{figure}[h]
\centering
\includegraphics[height=6cm, width=9cm]{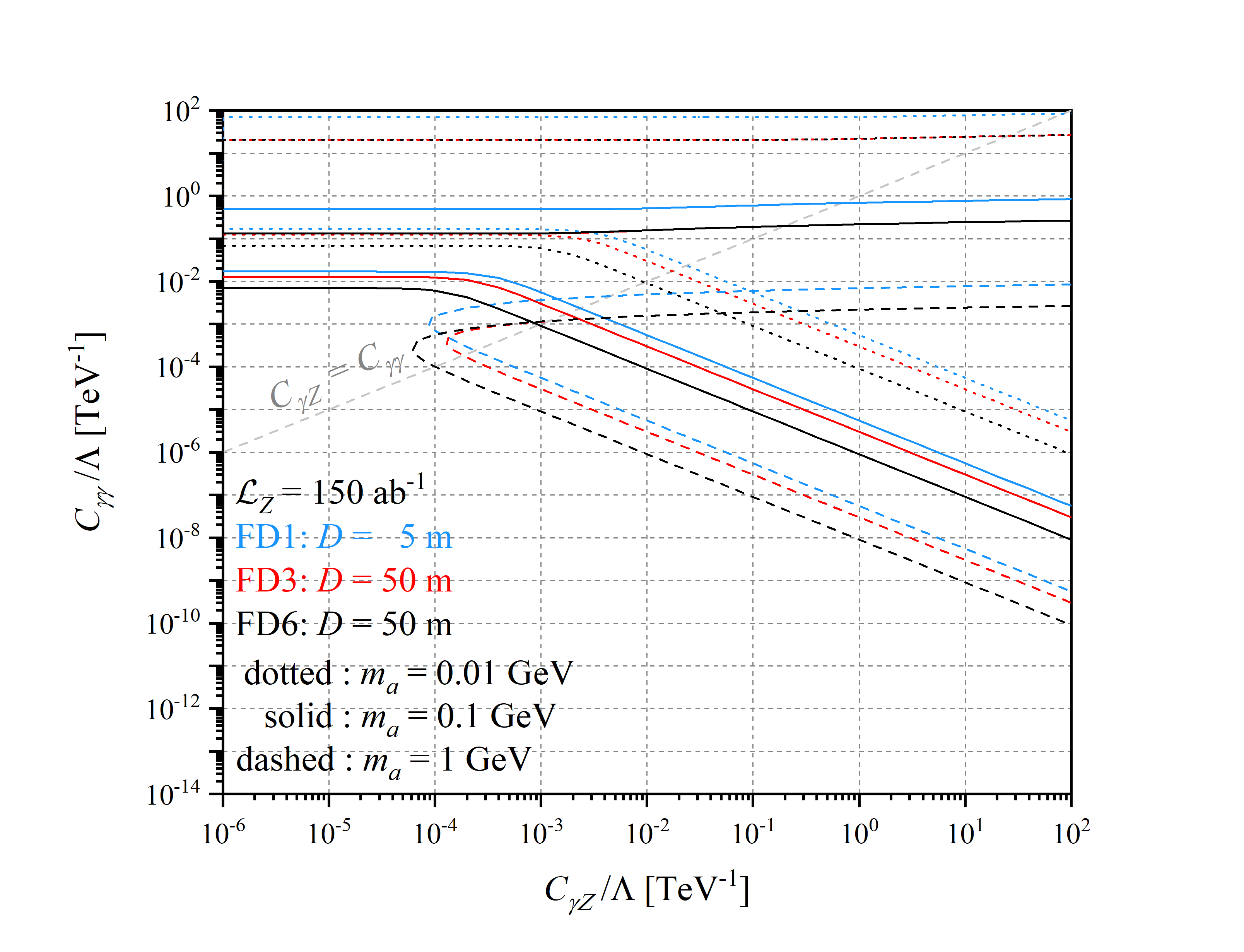}
\includegraphics[height=6cm, width=9cm]{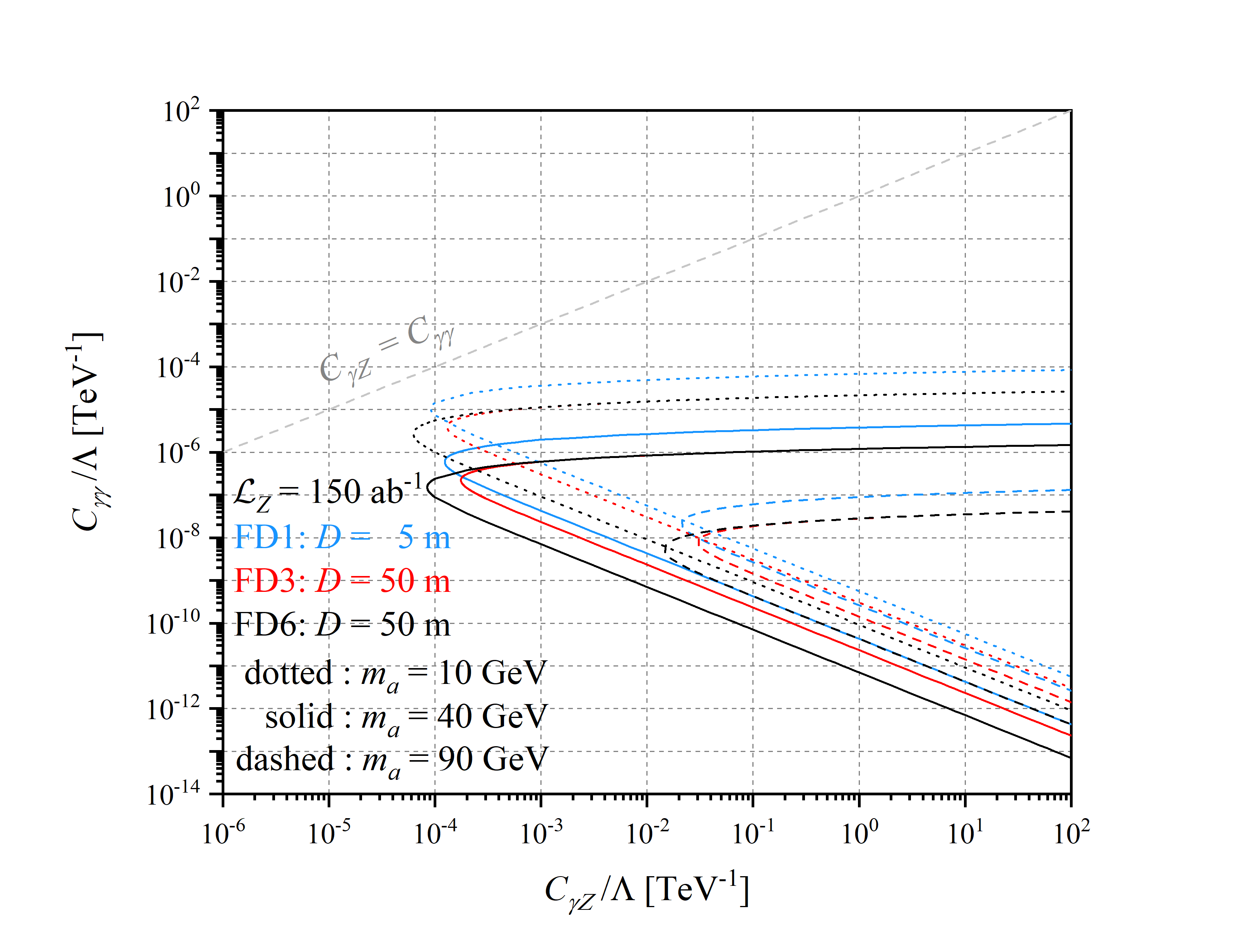}
\caption{
Upper: Sensitivity reaches of representative far detectors with $m_a =$ 0.01 GeV (dotted line), 0.1 GeV (solid line), and 1 GeV (dashed line) in the $C_{\gamma\gamma}/\Lambda$ vs $C_{\gamma Z}/\Lambda$ plane when integrated luminosity is $\mathcal{L}_Z$ = 150 ab$^{-1}$ . 
Lower: Sensitivity reaches of far detectors with $m_a = 10$ GeV (solid line), 40 GeV (dotted line), and 90 GeV (dashed line).
}
\label{fig:limitsM10}
\end{figure}

The limits are presented in the $C_{\gamma \gamma}/\Lambda$ vs. $C_{\gamma Z}/\Lambda$ plane in Fig.~\ref{fig:limitsM1} for benchmark value of $m_a$ = 1 GeV. 
The upper plot shows the sensitivity reaches of the representative far detectors with both options of $D$ and the integrated luminosity of $\mathcal{L}_Z$ = 150 ab$^{-1}$, 
while the lower plot shows the limits with various integrated luminosities of $\mathcal{L}_Z = 16, 150$ and 750 $\iab$. In this figure and Fig.~\ref{fig:limitsM10}, the gray dashed line represents the special case with $C_{\gamma Z} = C_{\gamma\gamma}$. 
It is obvious that the shapes of the curves are quite different from those in Figs.~\ref{fig:limitsCgz0} and~\ref{fig:limitsCeq}. In the upper curve, FD6 has the lowest $C_{\gamma Z}/\Lambda$ reach to $6.2\times10^{-5}~\iTeV$ with $C_{\gamma\gamma}/\Lambda = 2.5\times10^{-4}~\iTeV$; the $C_{\gamma Z}/\Lambda$ reaches of FD1 and FD5 are the same to $8.7\times10^{-5}~\iTeV$ with $C_{\gamma\gamma}/\Lambda = 1.1\times10^{-3}~ \text{and}~ 3.6\times10^{-4}~\iTeV$, respectively. The sensitivities of FD3 and FD8 are weaker compared with the other detectors. In the lower plot, the FD1's limits on $C_{\gamma Z}/\Lambda$ can reach as low as $3.0\times10^{-4}, 8.7\times10^{-5}, 3.4\times10^{-5}~\iTeV $ for 16, 150, 750 $\iab$ luminosities, respectively, while the lowest limits on $C_{\gamma Z}/\Lambda$ are $4.0\times10^{-4}, 1.3\times10^{-4}, 5.7\times10^{-5}~\iTeV$ and $1.9\times10^{-4}, 6.2\times10^{-5}, 2.8\times10^{-5}~\iTeV$ for FD3 and FD6, respectively. 
Similar to the figures in the previous two cases, compared with other detectors, FD1 is still competitive in the upper region with higher $C_{\gamma\gamma} / \Lambda$, while FD6 can probe smaller $C_{\gamma\gamma} / \Lambda$ in the lower region. 

In Fig~\ref{fig:limitsM10}, we compare the performances of the FD1, FD3, and FD6 with various mass values. We show results in two plots. The upper plot presents discovery regions for $m_a =$ 0.01 GeV (dotted line), 0.1 GeV (solid line), and 1 GeV (dashed line) in the $C_{\gamma\gamma}/\Lambda$ vs $C_{\gamma Z}/\Lambda$ plane with integrated luminosity of $\mathcal{L}_Z$ = 150 ab$^{-1}$, while the lower plot shows the discoverable parameter space for $m_a =$ 10 GeV (dotted line), 40 GeV (solid line), and 90 GeV (dashed line).

The behaviors of the contour curves can be understood from Eqs.~(\ref{eqn:crs}), (\ref{Eq:calSignal}), (\ref{eqn:lambda}) and Fig.~\ref{fig:ADPvsLambda} such that the signal rate is proportional to the production cross section times the average decay probability, i.e. $N_{\rm{ALP}}^{\rm{obs}}  \propto  \sigma(e^- e^+ \rightarrow \gamma a) \times \langle P[\rm{ALP\,in\,f.v.}]\rangle$, which is a complicated function of model parameters $C_{\gamma\gamma}/\Lambda, C_{\gamma Z}/\Lambda$, and $m_a$ with the following characteristics: 
(i) when $m_a \ll \sqrt{s}$, $(1 - m_a^2 / s)^3 \sim 1$ and $\sigma$ is insensitive to $m_a$; 
(ii) for fixed $m_a$, when $C_{\gamma Z} / \Lambda \gtrsim 2\times10^{-2}\, C_{\gamma \gamma} / \Lambda$,  the production cross section $\sigma$ is mainly determined by $C_{\gamma Z} / \Lambda$ rather than $C_{\gamma \gamma} / \Lambda$, and vice versa; 
(iii)  as shown in Fig.~\ref{fig:ADPvsLambda}, the average decay probability $\langle P[\rm{ALP\,in\,f.v.}]\rangle$ is tiny for extreme long decay length $\lambda$, and increases to the peak value and then decreases as $\lambda$ continously decreases;
(iv)  the decay length $\lambda$ is proportional to $(C_{\gamma\gamma} / \Lambda)^{-2}$, insensitive to $C_{\gamma Z} / \Lambda$, and proportional to $m_a^{-4}$ when $m_a \ll \sqrt{s}$.

For example, the discovery region for $m_a =$ 1 GeV satisfies $C_{\gamma Z} / \Lambda \gtrsim 2\times10^{-2}\, C_{\gamma \gamma} / \Lambda$, so $\sigma$ is mainly determined by $C_{\gamma Z} / \Lambda$ and insensitive to  $C_{\gamma \gamma} / \Lambda$. 
To understand the shape of the limit curves, one can start at the bottom right point and view the curves clockwise. 
The bottom right point with $C_{\gamma\gamma}/\Lambda \sim 10^{-10}\,\, \iTeV$ and $C_{\gamma Z}/\Lambda \sim 10^{2}\,\, \iTeV$ corresponds to the bottom right corner of the curve in Fig.~\ref{fig:ADPvsLambda} with $\lambda \sim 10^{15}$ m. As $C_{\gamma Z} / \Lambda$ decreases, $\sigma$ also decreases. To stabilize the signal rate, the average decay probability needs to be increased corresponding to decreasing of $\lambda$ which explains the increasing of $C_{\gamma \gamma} / \Lambda$ in the lower part of the limit curve.
The left extreme point with $C_{\gamma Z}/\Lambda \sim 10^{-4}\,\, \iTeV$ corresponds the peak position of the curve in Fig.~\ref{fig:ADPvsLambda}. After this point, as $C_{\gamma \gamma}/\Lambda$ increases, $\lambda$ is sequentially decreased across the peak point in Fig.~\ref{fig:ADPvsLambda} and thus the average decay probability decreases. Again, to stabilize the signal rate, $\sigma$ needs to be increased which explains the increasing of $C_{\gamma Z} / \Lambda$ in the upper part of the limit curve.
To summarize, the shapes of the limit curves for $m_a =$ 1 GeV (and also for $m_a =$ 10, 40, 90 GeV) can be understood by the counterbalance between $\sigma$ which is determined mainly by $C_{\gamma Z} / \Lambda$  and the average decay probability which depends on $\lambda$ and is affected more by $C_{\gamma \gamma} / \Lambda$.

The shapes of the limit curves for $m_a =$ 0.1 and 0.01 GeV are different from that for $m_a = 1$ GeV.
The lower boundary curve after $C_{\gamma Z} / \Lambda \lesssim 2\times 10^{-4}\,\, (2\times 10^{-3} )\,\, \iTeV$ and the whole upper boundary curve for $m_a =$ 0.1 (0.01) GeV become flat meaning they are insensitive to  $C_{\gamma Z} / \Lambda$. This is because after this turning point, $C_{\gamma Z} / \Lambda \lesssim 2\times10^{-2}\, C_{\gamma \gamma} / \Lambda$, from Eq.~(\ref{eqn:crs}), $\sigma$ is mainly determined by $C_{\gamma \gamma} / \Lambda$, and becomes insensitive to $C_{\gamma Z} / \Lambda$. Since $\lambda$ and hence the average decay probability do not depend on $C_{\gamma Z} / \Lambda$ either, the signal rate becomes insensitive to $C_{\gamma Z} / \Lambda$ after the turning point.

The discoverable regions shift downward as $m_a$ increases. 
This is mainly because, as shown in Eq.~\ref{eqn:lambda} $\lambda$ decreases rapidly with increasing $m_a$, and to maintain the $\lambda$ value $C_{\gamma\gamma}/\Lambda$ needs to be reduced accordingly.
Particularly, when $m_a \ll \sqrt{s}$, $\lambda \propto m_a^{-4}\, (C_{\gamma\gamma} / \Lambda)^{-2}$.
In such cases,  when $m_a$ increases by a factor of 10, $C_{\gamma\gamma}/\Lambda$ needs to be reduced by a factor of 100.
Comparing the discoverable regions for $m_a = 0.01, 0.1, 1 ,10$ GeV, one sees the downward shifts with a magnitude of $\sim 100$ correspondingly.
Moreover, comparing $m_a =$ 40 and 90 GeV, the discovery regions shift rightward. 
This is because as shown in Eq.~(\ref{eqn:crs}), when $m_a \sim \sqrt{s}$, $\sigma$ decreases obviously with increasing $m_a$.
Thus, to maintain the $\sigma$ value, $C_{\gamma Z}/\Lambda$ needs to increase.

\section{Conclusion and Discussion}
\label{sec:conc}

At generic future high energy $e^- e^+$ colliders, new detectors can be installed at a position far from the IP.
The LLPs produced at the IP can travel a long distance and decay inside the far detectors. Therefore, in principle such new experiments can enhance the discovery potential of LLPs.
In order to optimize the detector design, it is important to investigate sensitivities of different FD designs to various signals with typical production and decay modes.
We present a search strategy for the long-lived ALPs and explore the discovery sensitivities of eight different far detectors with different locations, volumes and geometries at future lepton colliders such as CEPC and FCC-ee.
We focus on the ALP couplings to the SM electroweak gauge bosons.
The ALPs are considered to be produced in association with one photon and decay into two photons, i.e. the signal process $e^-e^+ \rightarrow \gamma \,\, a,~ a \to \gamma\gamma $.

To exploit the high luminosities, signal events are simulated at center-of-mass energy of $\sqrt{s} = 91.2$ GeV.
We plot the polar angle distribution of ALPs and find that it is insensitive to the ALP masses and has two peaks around $90^\circ \pm 40^\circ$. Thus, most ALPs travel transversely, which means to maximize the acceptance of the considered signal, far detectors should be installed at the direction perpendicular to the collider beam and detectors located at the very forward direction downstream of the IP are disfavored.

Since the signal rate is proportional to the average decay probability $\langle P[\text{ALP}\text{ in f.v.}]\rangle$ of the ALPs inside the detector's fiducial volume, the average decay probabilities can affect the discovery sensitivities greatly.
In Fig.~\ref{fig:ADPvsLambda}, we present the average decay probability of produced ALPs in different far detectors as a function of  the decay length $\lambda$ in the laboratory frame, and find that far detectors with smaller distance from the IP have higher probabilities, which means closer distance is helpful to improve the discovery potentials.

We estimate the FDs' discovery sensitivities of long-lived ALPs for three physics scenarios:  $C_{\gamma Z} = 0$; $ C_{\gamma Z} = C_{\gamma \gamma}$ and both $C_{\gamma Z}$ and $C_{\gamma \gamma}$ can freely change.
For all three cases, the far detectors with smaller $D$ can give stronger discovery limits.
To compare all detectors, in general, 
FD1 is competitive in the upper parameter region with higher $C_{\gamma \gamma} / \Lambda$, while FD6 can probe smaller $C_{\gamma \gamma} / \Lambda$ in the lower region.
The performance of FD4 is almost identical to FD3, and the sensitivity of FD5 is between those of FD3 and FD6. 
The performance of FD7 is slight weaker than FD8. 
In general, among all far-detector designs, since FD1 is closer to the IPs and FD6 has bigger volume, they are expected to have the strongest discovery potentials. FD8 has weaker discovery potential than FD3, which means that increasing the length in the beam direction can not increase its discovery potential for the ALP signal.
Therefore, closer distance from the IP, bigger volumes, and the location lying at the direction perpendicular to the collider beam are proved to be useful to improve the discovery potential for the long-lived ALP signal based on this study.

When $C_{\gamma Z} = 0$ and integrated luminosity of $\mathcal{L}_Z$ = 150 ab$^{-1}$, FD1 has the largest mass reach to $m_a  = 0.54 $ GeV with $C_{\gamma\gamma}/\Lambda = 5.5\times10^{-3}~\iTeV$. 
The mass reaches of FD3 and FD6 can be $\sim$ 0.2\,–\,0.4 GeV with $C_{\gamma\gamma}/\Lambda$ = 7.0$\times10^{-3}$ and 4.0$\times10^{-3}~\iTeV$, respectively. 
When $C_{\gamma Z} = C_{\gamma \gamma}$,  because $C_{\gamma Z}$ can greatly enhance the signal production cross sections, the shapes of the discoverable parameter regions are similar to those of  $C_{\gamma Z} = 0$, but the boundary limits can be expanded.
With $\mathcal{L}_Z$ = 150 ab$^{-1}$, FD1 has the largest mass reach to $m_a =$ 4 GeV with $C_{\gamma\gamma}/\Lambda = 1.2\times10^{-4}~\iTeV$. 
The mass reaches of FD3 and FD6 can be $\sim$ 2\,–\,2.4 GeV with $C_{\gamma\gamma}/\Lambda$ = 1.5$\times10^{-4}$ and 8.0$\times10^{-5}~\iTeV$, respectively. 

When $C_{\gamma Z}$ and $C_{\gamma \gamma}$ can freely change, we present the limits in the $C_{\gamma\gamma}/\Lambda$ vs $C_{\gamma Z}/\Lambda$ plane for fixed ALP masses.
We find that the discoverable regions shift downward with increasing $m_a$.
For $m_a =$ 1 GeV and integrated luminosity of $\mathcal{L}_Z$ = 150 ab$^{-1}$, FD6 has the lowest $C_{\gamma Z}/\Lambda$ reach to $6.2\times10^{-5}~\iTeV$ with $C_{\gamma\gamma}/\Lambda = 2.5\times10^{-4}~\iTeV$.
We also observe that the discovery regions shift rightward when $m_a >$ 40 GeV. 
For $\mathcal{L}_Z$ = 150 ab$^{-1}$ and $m_a =$ 40 (90) GeV, FD6 has the lowest $C_{\gamma Z}/\Lambda$ reach to $8.5\times10^{-5}\,\,(1.5\times10^{-2})~\iTeV$ with $C_{\gamma\gamma}/\Lambda = 1.5\times10^{-7}\,\, (5.5\times10^{-9})~\iTeV$, respectively.

To estimate the effects of integrated luminosities, we compare the sensitivities for different luminosities of 16, 150, and 750 $\iab$ and find that larger luminosity is helpful to probe more parameter space with lower $C_{\gamma\gamma}/\Lambda$, while it cannot extend the upper side of the limit boundary of the parameter space with large $C_{\gamma\gamma}/\Lambda$ values.

We note that 
at electron-positron colliders, when assuming ALPs interact with $\gamma \gamma$, $\gamma Z$ and $Z Z$, they can also be produced from the vector boson fusion process $e^- e^+ \to e^- e^+ a$. The final state has an electron-positron pair and one ALP.
Since the associated particles are an electron-positron pair and visible in the near detector, in principle, the signal events accepted by the far detector can be added to enhance the discovery potential. However, on the one hand, because the kinematics, especially the distributions of the polar angle and speed, of ALPs are different between the $e^- e^+ \to e^- e^+ a$ and $\gamma a$ processes, the average decay probability of ALPs inside the far detector is also different. On the other hand, the production cross sections also have different dependences on the model parameters. Therefore, careful studies are needed to derive reliable enhanced limits.

Furthermore, from Fig.~\ref{fig:ADPvsLambda} and our sensitivity results, FD1 which is closest to the IP can be the most sensitive to the parameter regions with small decay length $\lambda$. 
Since the near detector is even closer than FD1, in principle, it is quite possible that the near detector could have more discovery potential to the parameter space with smaller decay length compared with FDs.
When the decay length of an ALP is around 5 m,  sizable number of signal events might be accepted in the near detector. The signal can be probed via two different final states. 
On the one hand, when ALPs are considered to decay inside the near detector, the final state has one photon from IP and two additional photons from the long-lived ALP decay. The displaced vertex needs to be reconstructed  to reduce the SM background. However, because the measurement of the directions of final state photons are rather difficult in the near detector, it is challenging to reconstruct the displaced vertex.
On the other hand, when ALPs are considered to decay outside the near detector, the final state has one photon from IP and large missing energy, which can also have large background from the SM process $e^- e^+ \to \gamma\, \nu\, \bar{\nu}$, where $\nu$ and $\bar{\nu}$ are SM neutrinos and antineutrinos.
Therefore, for these two final states, both the signal and the corresponding SM background need to be analyzed carefully to derive realistic limits, and large background could lead to weak limits  at the near detector.
They are beyond the scope of this paper, and we leave them for future studies.

\begin{acknowledgments}
\noindent
We thank Lingxiao Bai, Filmon Andom Ghebretinsae, Haiyong Gu, Yu Gao, Ying-nan Mao, Manqi Ruan, Lian-Tao Wang, Huaqiao Zhang and other members in CEPC study group for useful discussions.
M.T. and K.W. are supported by the National Natural Science Foundation of China under grant no.~11905162, the Excellent Young Talents Program of the Wuhan University of Technology under grant no.~40122102, and the research program of the Wuhan University of Technology under grant no.~2020IB024.
Z.S.W. is supported by the Ministry of Science and Technology (MoST) of Taiwan with grant numbers MoST-109-2811-M-007-509 and MoST-110-2811-M-007-542-MY3.
The simulation and analysis work of this paper was completed with the computational cluster provided by the Theoretical Physics Group at the Department of Physics, School of Sciences, Wuhan University of Technology.
\end{acknowledgments}



\appendix

\bibliography{Refs}

\bibliographystyle{h-physrev5}

\end{document}